\documentclass[a4paper,twocolumn]{article}

\usepackage{graphicx}
\usepackage{amssymb}
\usepackage{amsmath}
\usepackage{amsthm}

\theoremstyle{plain}
\newtheorem{thm}{Theorem}[section]
\newtheorem{lem}[thm]{Lemma}
\newtheorem{prop}[thm]{Proposition}
\newtheorem{dfn}[thm]{Definition}
\newtheorem{RMK}[thm]{Remark}

\makeatletter
\newenvironment{nodotproof}[1][\proofname]{\par
\pushQED{\qed}%
\normalfont \topsep6\p@\@plus6\p@\relax
\trivlist
\item[\hskip\labelsep
\itshape
#1]\ignorespaces
}{%
\popQED\endtrivlist\@endpefalse
}
\makeatother

\makeatletter
\@addtoreset{equation}{section}
\def\theequation{\thesection.\arabic{equation}}
\makeatother

\makeatletter
\let\@@thanks\@empty
\def\thanks#1{%
   \footnotemark
   \protected@xdef\@@thanks{%
     \@@thanks
     \protect\footnotetext[\the\c@footnote]{#1}}}
\def\@thanks{%
   \if@twocolumn
      \ifnum\col@number=\@ne \global\let\@outer@thanks\@@thanks
      \else                  \@@thanks
      \fi
   \else
      \@@thanks
   \fi}
\global\let\@outer@thanks\@empty
\let\@saved@topnewpage\@topnewpage
\long\def\@topnewpage[#1]{%
   \@saved@topnewpage[#1]%
   \begingroup
   \def\thefootnote{\fnsymbol{footnote}}
   \@outer@thanks
   \endgroup
   \global\let\@outer@thanks\@empty}
\makeatother

\newcommand{\trace}{\mathrm{Tr}}
\newcommand{\E}{\mathrm{E}}
\newcommand{\brho}{\bar{\rho}}
\newcommand{\rhot}{\rho_t}
\newcommand{\rhoe}{\rho_{\mathrm{e}}}
\newcommand{\berho}{{\bar{\rho}}^{\mathrm e}}

\title{Synthesis of Stabilizing Switched Controllers for $N$-Dimensional \\Quantum Angular Momentum Systems}
\author{Kyosuke Matsumoto\thanks{Department of Information Physics and Computing, Graduate School of Information Science and Technology, University of Tokyo, Hongo 7-3-1, Bunkyo-ku, Tokyo 113-0033, Japan, 
E-mail~:~\{matsumoto@hil.t, Koji\_Tsumura@ipc.i, Shinji\_Hara@ipc.i\}.u-tokyo.ac.jp } , Koji Tsumura$^*$, and Shinji Hara$^*$}
\date{}
\begin{document}
\twocolumn[
\maketitle
\paragraph{Abstract:}This paper provides a class of feedback controllers that guarantee global stability of quantum angular momentum systems. The systems are in general finite dimensions and the stability is around an assigned eigenstate of observables with a specific form. It is realized by employing the control law which was proposed by Mirrahimi \& van Handel. The class of stabilizing controllers is parameterized by a switching parameter and we show that the parameter between $0$ and $1/N$ assures the stability, where $N$ is the dimension of the quantum systems.
\vspace{1cm}
]
\section{Introduction}
Recently it has been suggested that the stabilization of quantum angular momentum systems is required to realize quantum information technologies 
\cite{Nielsen}. 
Feedback control laws proposed before Mirrahimi \& van Handel are applicable only to low dimensional quantum angular momentum systems \cite{ex2,ex1,yamamoto6}. However, the stabilization of higher dimensional angular momentum systems is required for realizing these technologies \cite{Nielsen}.

In these situations, Mirrahimi \& van Handel \cite{mirrahimi} have developed a control law which globally stabilizes general finite-dimensional quantum angular momentum systems around an assigned eigenstate of the observables. As far as we know, this is the first result which gives a method to control general finite-dimensional quantum systems. The purpose of this paper is to support the result of \cite{mirrahimi}. 

The control law in \cite{mirrahimi} is parameterized by a switching parameter $\gamma$. 
Mirrahimi \& van Handel \cite{mirrahimi} shows the existence of the switching parameter $\gamma$, which guarantees the stability of the quantum systems. 
From the view point of designing controllers, the result of \cite{mirrahimi} does not make clear what value of $\gamma$ the designer should set for controllers. 
This is our motivation in this paper, where we will explicitly show a set of $\gamma$ which guarantees the global stability. Thereby we can synthesis a stabilizing controller. 

The paper is organized as follows. 
In Section~2 we briefly review the quantum feedback control and introduce theorems 
used in this paper.
Section~3 formulates the stabilizing problem investigated in this paper and explains the result by Mirrahimi \& van Handel \cite{mirrahimi}.
Section~4 is devoted to the main result, where we provide a class of controllers which globally stabilize the system and guarantee the convergence to an assigned eigenstate. A numerical example is given to show the effectiveness of the main result in Section~5.  
In Section~6, we conclude the paper.

\section{Preliminaries}
In this section, we briefly review the quantum feedback control and show two theorems used in Section~4. 
This section includes fundamental results about quantum mechanics and stochastic processes. Note that some of these results are explained with a specific form which is naturally applicable to quantum feedback control problems in this paper. 

\paragraph{Quantum state}
In this paper, we consider a quantum system with dimension $1< N < \infty$. 
The quantum state of the system is denoted by $\rho$, an operator in a Hilbert space, which belongs to the state space
\begin{align}
\mathcal{S} = \{ \rho \in \mathbb{C}^{N \times N}: \rho = \rho^* ,\ \trace\rho = 1,\ \rho \ge 0 \}, 
\end{align}
where $\rho^*$ denotes Hermitian conjugation of $\rho$ and $\mathbb{C}^{N \times N}$  is a set of all $N \times N$ complex matrices. 
\paragraph{Quantum measurement}
We consider the case of that observables are Hermite matrices in $\mathbb{C}^{N\times N}$. 
We also assume that an observable $A$ is not degenerated, that is $A$ has mutually different eigenvalues $\{ a_k \}$. 
In the case of \textit{orthogonal measurement} \cite{Merzbacher} with the observable $A$ of a quantum state $\rho$, 
the numerical outcome is randomly selected from $\{ a_k \} ( k = 1, \ldots, N) $. The outcome $a_k$ is observed with probability
\begin{align}
\mathrm{Prob}(a_k) := {\psi(a_k)}^* \rho \ \psi(a_k), 
\end{align}
where $\psi(a_k)$ denotes a corresponding normalized eigenvector of $A$.
This measurement causes a jump of the state $\rho$ to the \textit{eigenstate} of $A$, ${\psi(a_k)} {\psi(a_k)}^*$.
Because of this jump it is difficult to realize a quantum feedback control by an orthogonal measurement. This is a motivation for employing a continuous measurement \cite{continuous1,continuous2}.

\paragraph{Feedback control of quantum systems}
In this paper, we consider quantum controlled systems via a continuous measurement. 
The dynamics of such a system with feedback obeys the following It$\hat{\mathrm{o}}$ equation for the conditional density $\rho_t$, which is a quantum analog of the Kushner-Stratonovich equation of nonlinear filtering \cite{continuous1,SSE,reduction}:
\begin{align}\label{originalmaster}
d\rhot = &-i[H,\rhot]dt -i u_t[G, \rhot]dt\notag\\ 
		 &+ (c \rhot c^* - \frac{1}{2}(c^*c\rhot + \rhot c^* c))dt \notag\\
         &+ \sqrt{\eta} ( c\rhot + \rhot c^* - \trace[(c+c^*)\rhot]\rhot)dW_t,
\end{align}
where
\begin{align}
 \rho_t: &\text{ a quantum state at time }t, \notag\\
 \rho_0: &\text{ an initial state}, \notag\\
 dW_t:   &\text{ an infinitesimal Wiener increment satisfying}\notag\\
           &\text{ the \textit{It$\hat{o}$ rules}} : \E[(dW_t)^2] = dt, \E[dW_t] = 0,\notag\\
 H:      &\text{ system's intrinsic Hamiltonian},\notag\\
 G:      &\text{ a control Hamiltonian},\notag\\
 u_t:    &\text{ control input }(u_t \in \mathbb{R}),\notag\\
 c:      &\text{ an observable}\notag\\
 \eta:   &\text{ the detector efficiency } (0 < \eta \le 1).\notag
\end{align}
The third and fourth terms in \eqref{originalmaster} represent the deterministic and stochastic back-action of the measurement, respectively. It should be noted that the solution of \eqref{originalmaster} is continuous in time \cite{Oksendal2}.

We will investigate the control problem of feedback stabilization of the equation.

\paragraph{Stability of the quantum system}
The solution of the equation \eqref{originalmaster} is a Markov process.
We define the stochastic stability of quantum systems as below by following Kushner \cite{Kushner1}:
\begin{dfn}
Let $\rhoe$ be an equilibrium point of \eqref{originalmaster}, i.e. $d\rho_t|_{\rho_t=\rhoe} = 0$. 
\begin{enumerate}
\item The equilibrium $\rhoe$ is said to be \textit{stable in probability} if
\begin{align}
&
\forall \epsilon > 0, \;\; 
\exists r(\epsilon) : (0,\infty) \rightarrow \mathbb{R}_{+}:=[0,\infty)
\nonumber
\\
&
\;\; \mbox{s.t.} \;\; \notag\\
&
\;\; \| \rho_0 - \rhoe \| < r(\epsilon) \notag\\
&
\;\;\;\; \Rightarrow \Pr \left(\sup_{0 \le t < \infty} \| \rho_t - \rhoe \| \ge \epsilon \right) = 0, 
\end{align}
where $\| \cdot \|$ is an arbitrary norm of a matrix in $\mathbb{C}^{N \times N}$. 
\item The equilibrium $\rhoe$ is \textit{globally stable} if it is stable in probability and additionally
\begin{align}
\forall \rho_0 \in \mathcal{S} \quad \Pr \left( \lim_{t \rightarrow \infty} \rho_t = \rhoe  \right) = 1.
\end{align}
\end{enumerate}
\end{dfn}
\paragraph{Preliminary results}
Before closing this section, we introduce two theorems which will be used in the proofs in Section~4. 

We can apply a theorem in \cite{Dynkin} pp.~111, Lemma~4.3 to \eqref{originalmaster} from the viewpoint that \eqref{originalmaster} is a transition equation of Markov process. Thereby we obtain the following theorem:
\begin{thm}\textnormal{\cite{Dynkin}}\label{Dynkin}
Consider a diffusion process $\rho_t \in \mathcal{S}$ starting from $\rho_0$. Let $\Gamma$ be an open subset of $\mathcal{S}$ and $\tau_{\rho_0}(\Gamma)$ be the first exit time of $\rho_t$ from $\Gamma$. Then for all $T \ge 0, \rho_0 \in \mathcal{S}$,
\begin{align}
\E [\tau_{\rho_0}(\Gamma)] \le \frac{T}{1-\sup_{\rho_0\in \mathcal{S}}\Pr\{ \tau_{\rho_0}(\Gamma) > T \}} .
\end{align}
\end{thm}

The other theorem is applicable to state equations for deterministic systems which is investigated in Section~4. We quote the theorem from \cite{LaSalle}.
\begin{dfn}[Invariant set \cite{LaSalle}]
An \textit{invariant set} $C$ is defined as a set with the property that if the initial state of the system is in $C$ then its whole path (forward and backward) lies in $C$.
\end{dfn}

\begin{thm}[LaSalle's invariance principle \cite{LaSalle}]\label{LaSalle}
Let $Q(\rho):\mathcal{S}\rightarrow \mathbb{R}_{+}$ be a scalar function with continuous first partial derivatives. Let $\Omega_l$ designate the region where $Q(\rho) < l$. Assume that $\Omega_l$ is bounded and that within $\Omega_l$,  $Q(\rho)$ is positive definite and $\frac{d Q(\rho)}{d t} \le 0$. Let $R$ be the set of all points within $\Omega_l$ where $\frac{d Q(\rho)}{d t} = 0$, and let $C'$ be the largest invariant set in $R$. Then every solution $\rho_t$ in $\Omega_l$ tends to $C'$ as $t \rightarrow \infty$
\end{thm}

\section{Control law}

The purpose of this section is to introduce the control law given by
Mirrahimi \& van Handel \cite{mirrahimi}.

\paragraph{Stabilizing control problem}

At first we explain the control problem of the quantum systems \eqref{originalmaster} as follows.  
The quantum systems are assumed to be under continuous measurement of a
fixed angular momentum $J$ ( $J$ is a positive integer or half-integer ).  This means the dimension $N$
of the systems is $2J+1$.  A typical case of such systems is an atomic
ensemble detected through a dispersive optical probe \cite{system}.  We measure the angular momentum of the systems along an axis, say $z$ and control the angular momentum by applying a magnetic field to the system along an axis, say $y$ which is in a direction perpendicular to an axis $z$. 
The control objective is to move the quantum state into an eigenstate of the angular
momentum. 

The dynamics of such quantum systems under an ideal condition can be
described by \eqref{originalmaster} with the three operators; $H$, $G$, $c$, defined as
follows:\\  
1. $H = 0$. 
\\
2. $G=\beta F_y$ ($\beta > 0$), where $F_y$ represents the angular
momentum along the axis $y$ of the form \cite{Merzbacher}
\begin{align}
\label{eq:Fy}
F_y&=\frac{1}{2i}
\begin{bmatrix}
0     & -c_{1}  &        &          &    O   \\
c_{1} & 0       & -c_{2} &          &        \\
      & \ddots  & \ddots & \ddots   &        \\
      &         &c_{N-2} & 0        &-c_{N-1}\\
O     &         &        & c_{N-1}  & 0     \\
\end{bmatrix},\notag\\
c_k&=\sqrt{(N-k)k} = \sqrt{(2J+1-k)k}.
\end{align}
\\
3. $c = \alpha F_z$ ($\alpha > 0$), where $F_z$ represents the angular
   momentum along the axis $z$ of the form \cite{Merzbacher}
\begin{align}
F_z=
\begin{bmatrix}
-J    &      &       &     & O   \\
      &-J+1  &       &     &   \\
      &      &\ddots &     &   \\
      &      &	     &J-1  &   \\
  O    &      &       &     &J  \\
\end{bmatrix}.
\end{align}
Let $\lambda_k := k-J-1\ (k = 1,\ldots,N)$ denotes an eigenvalue of $F_z$ and $\psi_k$ is the corresponding eigenvector. 

\noindent
Furthermore without loss of generality we assume $\alpha = \beta = 1$ by
scaling of the time and $u_t$ and we obtain: 
\begin{align}
d\rho_t = &-i u_t[F_y,\rho_t]dt -\frac{1}{2}[F_z,[F_z,\rho_t]]dt \notag\\
&+ \sqrt{\eta}(F_z \rhot +\rhot F_z - 2 \trace[F_z\rhot]\rhot)dW_t,  \label{master}
\end{align}
where $0 < \eta \le 1$. 
The control objective is to stabilize 
$\rho_{(f)} := \psi_f \psi_f^*$ globally where $f$ is one of indices $1,\ldots,N$. 
We define a distance function
\begin{align}
V(\rho) := 1 - \trace(\rho\rho_{(f)}) = 1 - (\rho)_{ff}: \mathcal{S}\rightarrow [0,1]
\end{align}
from the state $\rho$ to the target state $\rho_{(f)} $ for the explanation of the result of \cite{mirrahimi} and the proof of our main result. 

Mirrahimi \& van Handel \cite{mirrahimi} have proposed a control law for this stabilization problem as in the following theorem: 
\begin{thm}\textnormal{\cite{mirrahimi}}
\label{MH} Consider the system \eqref{master} evolving in the set $\mathcal{S}$. Let $\gamma > 0$. Consider the following control law:
\begin{enumerate}
\item $u_t = - \trace(i[F_y ,\rhot]\rho_{(f)})\ if\ V(\rhot ) \le 1- \gamma;$
\item $u_t = 1\ if\ V(\rhot ) \ge 1-\gamma/2;$
\item If $\rhot \in \mathcal{B} = \{\rho : 1-\gamma < V(\rho) <1-\gamma/2\}$, then $u_t = -\trace(i[F_y,\rhot]\rho_{(f)})$ if $\rhot$ last entered $\mathcal{B}$ through the boundary $V(\rhot) = 1-\gamma$, and $u_t = 1$ otherwise.
\end{enumerate}
Then $\exists \gamma >0$ s.t. $u_t$ globally stabilizes \eqref{master} around $\rho_{(f)} $ and $\E[\rhot] \rightarrow \rho_{(f)} $ as $t \rightarrow \infty$.
				 \end{thm}
Hereafter we call the control law in Theorem~\ref{MH} \textit{MH
 control law}  
and $u_\mathrm{MH}(\gamma)$ denotes the control input using MH control law with the parameter $\gamma$.

\section{Synthesis of stabilizing controllers}
\subsection{Main theorem}
Theorem~\ref{MH} shows the existence of a positive parameter $\gamma$
with which the stability of the quantum systems is guaranteed. 
In other words, Theorem~\ref{MH} is the existence theorem for the
stabilizing controllers.  On the other hand, from the viewpoint of 
controller design, it is unclear what positive number $\gamma$, which
the designer should set for the controllers, guarantees the stability.   
In this section, we give a set of $\gamma$, which definitely guarantees
the stability of the quantum systems, in the following theorem which is
a main result of this paper. 
\begin{thm}
\label{mythm}Consider the system \eqref{master} evolving in the set $\mathcal{S}$. Let $u_t = u_\mathrm{MH}(\gamma)$. If $\gamma \in (0, \frac{1}{N})$, $u_t$ globally stabilizes \eqref{master} around the target state $\rho_{(f)} $ and $\E[\rho_t] \rightarrow \rho_{(f)} $ as $t \rightarrow \infty$.
\end{thm}

The proof of Theorem~\ref{mythm} is executed in the following three steps:\\
1. At first, 
we show that the ensemble average of the solution of \eqref{master} with
 a $\rho_t$-independent nonzero control input $u_t \in C^1$ converges to \textit{the
      maximally mixed state}  as $t \rightarrow \infty$. \\
2. Secondly, we prove the following lemma, the proof of which is given in subsection~4.2:
\begin{lem}
\label{finite}
Consider the system \eqref{master} with $u_t = 1$ and an initial state $\rho_0 \in \mathcal{S}_{>1-\gamma_\mathrm{a}}:=\{ \rho \in \mathcal{S}: \gamma_\mathrm{a} < V(\rho)\le 1\} $. 
Let $\tau_{\rho_0}(\mathcal{S}_{>1-\gamma_\mathrm{a}})$ be the first exit time of $\rho$ from $\mathcal{S}_{>1-\gamma_\mathrm{a}}$. Then, if $0 <  \gamma_\mathrm{a} < 1/N$, 
\begin{align}\label{mygoal}
\sup_{\rho_0 \in \mathcal{S}_{>1-\gamma_\mathrm{a}}}E[\tau_{\rho_0}(\mathcal{S}_{>1-\gamma_\mathrm{a}})] < \infty. 
\end{align}
\end{lem}
\renewcommand{\qedsymbol}{$\Diamond$}
\begin{RMK}
\begin{nodotproof}[]
This lemma gives a set of $\gamma_\mathrm{a}$ which satisfies \eqref{mygoal}, 
while Mirrahimi \& van Handel \cite{mirrahimi} shows the existence of such $\gamma_\mathrm{a}$. 
This is a main defference between \cite{mirrahimi} and this paper. 
\end{nodotproof}
\end{RMK}
\renewcommand{\qedsymbol}{$\Box$}

\noindent
3. Finally, we complete the proof of Theorem \ref{mythm} by combining the results of the second step and the following lemma which we collect lemmas in \cite{mirrahimi} into:
\begin{lem}
\textnormal{\cite{mirrahimi}}
\label{MH2}
Consider the system \eqref{master} with $u_\mathrm{MH}(\gamma)$ and an initial state $\rho_0 \in \mathcal{S}$. If $\gamma = \gamma_\mathrm{a}$ satisfies $\eqref{mygoal}$,  $u_\mathrm{MH}(\gamma)$ globally stabilizes \eqref{master} around $\rho_{(f)} $ and $\E[\rhot] \rightarrow \rho_{(f)} $ as $t \rightarrow \infty$.
\end{lem}

\subsection{Proof of Theorem~\ref{mythm}}
This subsection is  devoted to the proof of Theorem~\ref{mythm}, which is devided into three parts corresponding to the three steps proposed in the previous subsection. 
\paragraph{Step 1.}
In this step, we consider the dynamics of $\E[\rho_t]=:\brho_t$, 
\begin{align}
d\brho_t = -iu_t[F_y,\brho_t]dt - \frac{1}{2}[F_z,[F_z,\brho_t]]dt\label{Emaster}, 
\end{align}
where a control input is nonzero and $\rho_t$-independent and has continuous first derivatives.
We call \eqref{Emaster} the ensemble dynamics.

At first we give the following lemma on the equilibrium points of \eqref{Emaster}. 
\begin{lem}
\label{equilibrium}
The ensemble dynamics \eqref{Emaster} has a unique equilibrium point 
$\displaystyle \frac{1}{N}I$, which is \textnormal{the maximally mixed
 state} of \eqref{originalmaster}.
\begin{proof}
Let us assume that the ensemble dynamics has an equilibrium point
 $\berho$. 
That is 
\begin{align}
\label{equi1}
-iu_t[F_y,\berho]dt -\frac{1}{2}[F_z,[F_z,\berho]]dt = 0. 
\end{align}
Multiplying both sides of \eqref{equi1} by $2\berho$ and subsequently taking trace of the both sides gives
\begin{align}
\label{equi11}
\trace([F_z,[F_z,\berho]]\berho)dt = 0.
\end{align}
Remember $F_z$ and $\berho$ are Hermitian matrices. 
The left side of \eqref{equi11} can be computed as 
\begin{align}
\label{equi2}
&\trace([F_z,[F_z,\berho]]\berho)dt  \notag\\
&= \trace\{(\berho)^* F_z^*[F_z, \berho] - [F_z,\berho] F_z^* (\berho)^*\}dt\notag\\
&= \trace\{(\berho)^* F_z^*[F_z, \berho] - F_z^* (\berho)^* [F_z,\berho] \}dt\notag\\
&= \trace([F_z,\berho]^*[F_z,\berho])dt\notag\\
&= \| [F_z,\berho] \|^2 dt, 
\end{align}
where $\| [F_z,\berho] \|$ denotes the \textit{Frobenius norm} of the matrix $[F_z,\berho]$. \eqref{equi1} and \eqref{equi2} show
\begin{align}\label{diagonal2}
([F_z,\berho])_{lm} &= 0\notag\\
& (l, m \in \mathbb{N},\ 1\le l, m \le N), 
\end{align}
where $([F_z,\berho])_{lm}$ stands for the $(l,m)$th element of the matrix $[F_z,\berho]$. 
Noting that $F_z$ is a diagonal matrix, we obtain
\begin{align}\label{diagonal}
&([F_z,\berho])_{lm} \notag\\
&=(\berho)_{lm}\{(F_z)_{ll} - (F_z)_{mm} \} =0, \notag\\
&\ \ \ \ \ \ \ \ \ \ \ \ \ (l, m \in \mathbb{N},\ 1\le l, m \le N).
\end{align}
Note also that $F_z$ has no repeated diagonal entries. By \eqref{diagonal},  $\berho$ should be a diagonal matrix.

Now we substitute a diagonal matrix $\brho_t = \berho$ into
\eqref{Emaster} and obtain
\begin{align}
\label{maximally}
&(-i[F_y,\berho])_{lm} \notag\\
&= -i(F_y)_{lm}\{(\berho)_{mm} -(\berho)_{ll} \} = 0, \notag\\
&\ \ \ \ \ \ \ \ \ \ \ \ \ (l,m \in \mathbb{N},\ 1\le l,m \le N).
\end{align}
Since the location of the nonzero elements of $F_y$ is limited as in \eqref{eq:Fy} and
 with \eqref{maximally}, 
the all diagonal elements of $\berho$ are known to be equal. 
Such a matrix in $\mathcal{S}$ is only \textit{the maximally mixed state} $\frac{1}{N}I$. This completes the proof.
\end{proof}
\end{lem}

We can derive the following proposition from this lemma. 
\begin{prop}
\label{converge}
Consider a $\rho_t$-independent nonzero control input $u_t \in C^1$. 
For any initial state $\rho_0 \in \mathcal{S}$, the solution $\brho_t$
 of  \eqref{Emaster} converges to
 \textit{the maximally mixed state} $\frac{1}{N}I$ as $t \rightarrow \infty$.
\begin{proof}
Consider a function
\begin{align}
Q(\brho_t) := \trace({\brho_t}^2) - \frac{1}{N} 
\end{align}
as a candidate of the Lyapunov function. 
It is easily verified with $\trace{\brho_t} = 1$ that $Q(\brho_t) \ge 0$ for all
 $\brho_t \in \mathcal{S}$ and that $Q(\brho_t) = 0$ iff
 $\brho_t = \frac{1}{N}I$. 
A computation similar to that shown in \eqref{equi1} and \eqref{equi2} gives 
\begin{align}\label{con1}
\frac{dQ(\brho_t)}{dt} = -\|[F_z,\brho_t]\|^2 \le 0.
\end{align}
We have equality in \eqref{con1} iff $\brho_t$ is a diagonal matrix (from the same argument in Lemma \ref{equilibrium}). Let $\mathcal{C'}$ be the invariant set contained in $\{\brho \in \mathcal{S}:\frac{dQ(\brho)}{dt} =0\}$. We can see the set $\mathcal{C'}$ includes any diagonal matrix. 

Recall the result and proof of Lemmma~\ref{equilibrium}. The maximally mixed state satisfies $\frac{dQ(\brho)}{dt}\Big|_{\brho=\frac{1}{N}I} =0 $ and it is an equilibrium point of the ensemble dynamics. So the maximally mixed state is an elemnt of $\mathcal{C'}$. 

On the other hand, let us assume that $\brho$ is a diagonal matrix other than the maximally mixed state at time $t$. 
We can obtain
\begin{align}\label{4.11}
(d\brho_t)_{lm} = -iu_t(F_y)_{lm}\{(\brho_t)_{mm}-(\brho_t)_{ll}\}, \notag\\ 
(l,m \in \mathbb{N}, 1\le l,m \le N, l \neq m, ). 
\end{align}
In the equation \eqref{4.11}, there exist $l$ and $m$ such that $(d\brho_t)_{lm}$ is not zero as we saw in the proof of Lemma~\ref{equilibrium}. That is, $\brho$ changes into an off-diagonal matrix. Therefore a diagonal matrix other than the maximally mixed state is not an element of $\mathcal{C'}$. 

Thus the maximally mixed state is the unique element of $\mathcal{C'}$. 
The assertion that $\brho_t \rightarrow  \frac{1}{N}I$ as $t \rightarrow \infty$ can be proved by applying Theorem=~\ref{LaSalle} (LaSalle's invariance principle). 
\end{proof}
\end{prop}
\paragraph{Step 2.}
In this step, 
we consider the system \eqref{master} with $u_t = 1$ and an initial
state $\rho_0 \in \mathcal{S}$. 
At first, we show the following lemma.
\begin{lem}
\label{V}
For $\gamma_\mathrm{a} \in (0,\frac{1}{N})$,
there exists a finite time $T_1$ such that 
\begin{align}
\min_{t\in[0, \; T_1]} E[V(\rho_t)] < 1-\gamma_\mathrm{a}.
\end{align}
\begin{proof}
The solution of \eqref{master} is continuous in $t$ and it is obvious that  $V(\rho_t)$ is continuous in $\rho_t$. Thus Proposition~\ref{converge} implies
\begin{align}
\lim_{t \rightarrow \infty}\E[V(\rho_t)] &= V\left(\lim_{t\rightarrow \infty}E[\rho_t]\right) \notag\\
&= 1-\frac{1}{N} < 1-\gamma_{\mathrm{a}}. 
\end{align}
That is,
\begin{align}
\label{V1}
&
\forall \epsilon > 0, \;\; 
\exists T_1
\nonumber
\\
&
\;\; \mbox{s.t.} \;\; t\ge T_1 \Rightarrow
 |E[V(\rho_t)]-(1-\frac{1}{N})| < \epsilon.
\end{align}
Therefore, in the case $\epsilon = \frac{1}{N} - \gamma_\mathrm{a}$, 
we obtain
\begin{align}
\exists T_1 < \infty  \;\; \mbox{s.t.} \;\; t\ge T_1 \Rightarrow E[V(\rho_t)] < 1 - \gamma_\mathrm{a}.
\end{align}
\end{proof}
\end{lem}

With this lemma, we can prove Lemma~\ref{finite}. 
\begin{proof}[Proof of Lemma~\ref{finite}]
By Lemma \ref{V}, 
\begin{align}
\label{VV}
\exists T_0 ;\; 
T_0 = \inf_{t\in [0, \; \infty)} \{t \; | \; \E[V(\rho_t)] \le 1-\gamma_{\mathrm{a}}\}. 
\end{align}
In addition, applying Theorem~\ref{Dynkin} to the case 
$\Gamma = \mathcal{S}_{>1-\gamma_\mathrm{a}}$ and $T = T_0$ 
yields to 
\begin{align}\label{4.17}
E[\tau_{\rho_0}(\mathcal{S}_{>1-\gamma_{\mathrm{a}}})] \le \frac{T_0}{1-\sup_{\zeta\in\mathcal{S}}\mathrm{Pr}\left\{\tau_\zeta(\mathcal{S}_{>1-\gamma_{\mathrm{a}}}) > T_0 \right\}}. 
\end{align}
Now we can prove this lemma in the same way with \eqref{VV}, \eqref{4.17} as the proof of Lemma~4.6 in Mirrahimi~\&~van~Handel \cite{mirrahimi}. 
\end{proof}

\paragraph{Step 3.}
\begin{proof}[Proof of Theorem~\ref{mythm}]
Consider the system \eqref{master} with the control input
 $u_\mathrm{MH}(\gamma)$ and an initial state $\rho_0 \in \mathcal{S}$.
When $\rho_0 \in \mathcal{S}_{> 1-\gamma}$, 
the control input is $u_t = 1$  until $\rho_t$ exits from $\mathcal{S}_{>1-\gamma}$. 
Therefore we can show that \eqref{mygoal} holds for
 $\gamma = \gamma_{\mathrm a} \in (0,\frac{1}{N})$ using Lemma \ref{finite}. 
Then, applying Lemma~\ref{MH2} completes the proof of Theorem~\ref{mythm}. 
\end{proof}

\section{Numerical example}
We here give a numerical example to show the effectiveness of the main result in this paper for synthesis of stabilizing controllers. We deal with a quantum system under a continuous measurement of a fixed angular momentum $J=10$, that is the dimension of the system $N$ is $21$. Let us stabilize \eqref{master} around $\rho_{(11)}$ globally from the initial state $\rho_{(1)}$ by utilizing the MH control law. 

\begin{figure}[htb]
\begin{center}
\includegraphics[width=7.2cm]{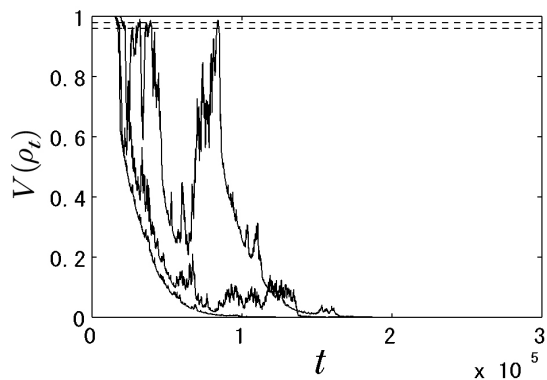}
\end{center}
\caption{In the case of $\gamma = 0.04 < \frac{1}{21}$, all the sample paths converge to $\rho_{(11)}$ from $\rho_{(1)}$. }\label{gamma004}
\end{figure}

\begin{figure}[htb]
\begin{center}
\includegraphics[width=7.2cm]{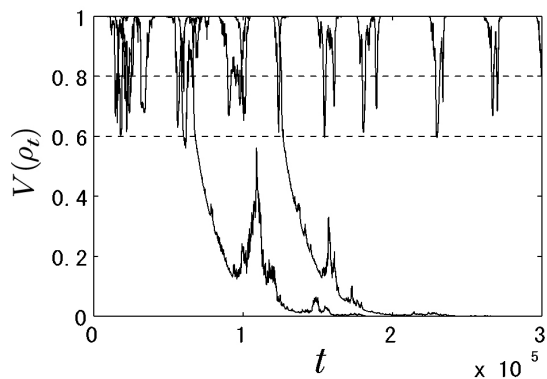}
\end{center}
\caption{In the case of $\gamma = 0.4 \ge \frac{1}{21}$ there exists a path which is expected not to cnverge to $\rho_{(11)}$. }\label{gamma04}
\end{figure}

We can see from Theorem~\ref{mythm} that we only have to set a switching parameter $\gamma$ between 0 to $\frac{1}{N}=\frac{1}{21}$ in order to realize the global stability. 
Figs.~{\ref{gamma004} and \ref{gamma04}} illustrate three sample paths 
of $V(\rho_t)$ in the case that $\gamma$ is set to $0.04 (<\frac{1}{21},$ Fig.~\ref{gamma004}$)$ and $0.4 (\ge\frac{1}{21},$ Fig.~\ref{gamma04}$)$ ,respectively. 
The value of $\gamma = 0.04$ guarantees the stability of the  21-dimensional quantum angular momentum system. However, the value of $\gamma = 0.4$ does not so. These numerical simulations illustrate the result of Theorem~\ref{mythm} and show the effectiveness of it.
\section{Conclusion}
This paper provides a class of controllers that guarantee the global
 stability of an assigned eigenstate for $N$-dimensional quantum angular momentum systems by utilizing the result of \cite{mirrahimi}. The main result of this paper is in Theorem \ref{mythm}, which says that any switching parameter $\gamma \in (0, \frac{1}{N})$ in a MH control law given by \cite{mirrahimi} is sufficient for
stability in probability 1. 

However, we should note that Theorem~\ref{mythm} gives only a sufficient condition for the stability in probability 1. A necessary and sufficient condition for the stability remains as a future interesting work.

\section*{Acksnowledgments}This research is supported in part by The Ministry of Education, Science, Sport and Culture, Japan, under Grant 17656137. 
\small

\end{document}